# Optical control of charged exciton states in tungsten disulfide

M. Currie[1], A.T. Hanbicki[1], G. Kioseoglou[2,3], and B.T. Jonker[1]

[1]*Naval Research Laboratory*, Washington, DC 20375
[2]*University of Crete*, Heraklion Crete, 71003, Greece
[3]*Institute of Electronic Structure and Laser (IESL), Foundation for Research and Technology Hellas (FORTH),* Heraklion Crete, 71110, Greece

ABSTRACT

A method is presented for *optically* preparing $WS_2$ monolayers to luminesce from only the charged exciton (trion) state--completely suppressing the neutral exciton. When isolating the trion state, we observed changes in the Raman $A_{1g}$ intensity and an enhanced feature on the low energy side of the $E^1_{2g}$ peak. Photoluminescence and optical reflectivity measurements confirm the existence of the prepared trion state. This technique also prepares intermediate regimes with controlled luminescence amplitudes of the neutral and charged exciton. This effect is reversible by exposing the sample to air, indicating the change is mitigated by surface interactions with the ambient environment. This method provides a tool to modify optical emission energy and to isolate physical processes in this and other two-dimensional materials.

Monolayer transition metal dichalcogenides (TMDs) are promising materials for future 2D nanoelectronic systems [1–3]. Their semiconductor band structure and relatively high mobilities enable monolayer transistors [4] and their direct gap enables such applications as optical sources and detectors [2]. The properties inherent in these 2D systems produce an all-surface material applicable to sensing[5], single-atom storage [6], quantum-state metamaterials [7], and other quantum-based technologies.

When $MX_2$ TMDs (where M = Mo, W and X = S, Se) are optically pumped, the resultant photoluminescence (PL) is characterized by excitonic emission. In undoped material, the main emission feature is from the neutral exciton, $X^0$. Charged excitons, or trions, become important in doped systems and have emission energies 30-50 meV below the neutral exciton [8]. The dynamic behavior of photoexcited carriers in $MX_2$ materials are governed by strong Coulomb interactions. These interactions have a profound influence on the charge concentration which in turn influences physical properties. Perturbations of properties such as conductivity, for instance, enable applications that probe charge-density changes such as the detection of DNA [9]. Modification of the carrier concentration in $MoS_2$, has been achieved by applying a gate bias (in FET devices) [8], physisorption [10], chemical doping [11], and functionalization [12]. Changes in carrier density can enhance the formation of charged excitons (trions) and manipulation of trions in both $MoS_2$ and $MoSe_2$ using electrostatic gating has been demonstrated [8,13].

There are a myriad of ways in which changes to the carrier concentration in $MX_2$ materials influences the optical properties, and more specifically the PL. For example, in $MoS_2$, increased electron doping suppresses optical absorption near the exciton peaks resulting in diminished exciton PL intensity. Concurrent with this intensity reduction is the emergence of a trion peak [8]. Conversely, $MoS_2$, which is typically n-type, can be transformed to heavily p-type via chemical absorption of oxygen at defect sites formed during annealing. This process converts negatively charged trions back to neutral excitons, resulting in an enhancement of the neutral exciton PL [14].

Similarly, physisorbed molecules (e.g., $O_2$ and $H_2O$) were shown to influence the PL in $MoS_2$, $MoSe_2$, and $WSe_2$ [10]. Annealing produced a large increase in the PL intensity for $MoS_2$ and $MoSe_2$ samples in $O_2$ and $H_2O$ environments, while PL decreased for $WSe_2$ samples. These changes are reversible and result from carrier transfer between the gas and the 2D material [10]. Finally, large PL intensity was observed from induced defects (specifically $MoS_2$, $MoSe_2$, and $WSe_2$) [10], where point defects were created by either high-temperature annealing or by alpha-particle irradiation[15]. The point defects consist of anion vacancies which in turn enable bound exciton formation. Gas molecules that interact with defect sites can enhance PL by draining electrons, shifting charged excitons to neutral excitons.

While much attention has been given to $MoS_2$, $MoSe_2$, and $WSe_2$, there has been a growing interest in $WS_2$ due to its large spin-orbit splitting in the valence band (420 meV) [16], large exciton/trion binding energy [17,18], and high emission quantum yield even at room temperature [19]. In addition to these intriguing optical properties, improved electrical performance of a single-layer $WS_2$ transistor was also observed after annealing samples at 115°C for 10 to 145 hours in vacuum [20]. In this letter, we demonstrate the ability to *optically* prepare $WS_2$ monolayers such that the emission is from either only the neutral exciton or only the charged exciton (trion) state. We demonstrate that this effect is completely reversible and is due to the surface condition of the monolayer. The surface sensitivity we observe is likely responsible for many emission characteristics seen in all TMDs, since these materials are dominated by their surface properties. Reliably and reproducibly controlling TMD absorption and emission behavior is important for physical measurements (e.g., time-resolved measurements of the pure exciton/trion state) as well as for most applications of these materials.

Monolayer $WS_2$ samples were prepared by mechanical exfoliation of bulk $WS_2$ onto $SiO_2$-on-Si substrates. Optical inspection, Raman spectroscopy, and atomic force microscopy were used to identify monolayer sample regions. One such exfoliated sample is shown in Fig. 1(a) where a >20-$\mu m^2$ region is identified as a monolayer, bordered by either substrate or multilayer regions. A micro-PL setup with a 50x objective and incorporating a continuous-flow He-cryostat was used to collect the PL and reflectivity in a backscattering geometry. Samples were excited with continuous-wave lasers of various wavelengths. Emitted light was dispersed by a single monochromator equipped with a multichannel charge coupled device (CCD) detector. The size of the optical beam we used for reflectivity and PL measurements is ~1µm in diameter as illustrated by the dotted circle superimposed on the monolayer region in Fig. 1(a).

As expected, strong photoluminescence is observed over the monolayer region while much weaker PL (by orders of magnitude) is observed in the multilayer regions. The PL at room temperature in ambient atmospheric conditions shows a single peak at 1.99 eV. The PL at room temperature in vacuum, after exposing the sample to the excitation laser for a few minutes shows a single peak at 1.94 eV. These two extremes are shown in Fig. 1(b) using a 594-nm excitation source at a power of 1 and 50 µW (respectively). The higher energy peak at 1.99 eV we attribute to the neutral exciton, and the peak at 1.94 eV to the trion. To quantify and understand this effect we systematically varied the optical excitation power while monitoring the PL as a function of environment and temperature to establish the evolution of the isolated trion state.

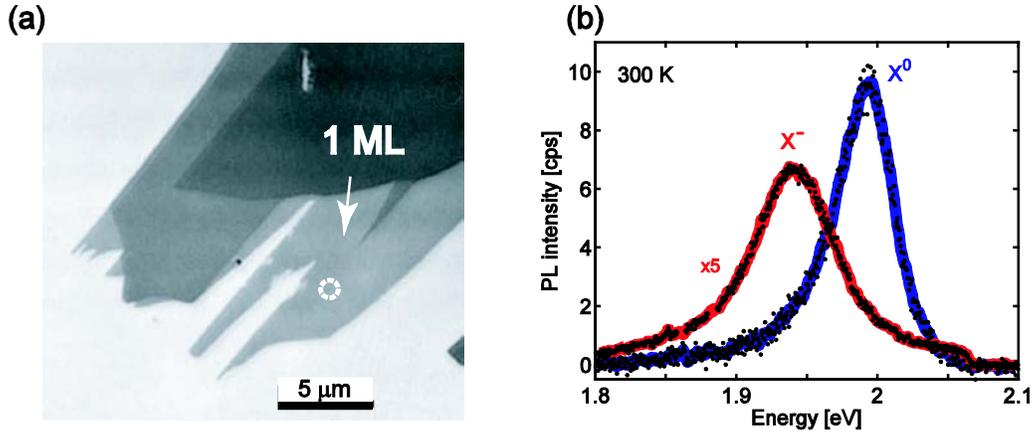

*Figure 1(a) Confocal microscope image of a representative exfoliated WS$_2$ sample. Arrow indicates monolayer region and dotted circle indicates approximate optical probe size. (b) Photoluminescence spectra of exfoliated sample taken with a 594 nm laser excitation at 300 K showing the neutral exciton ($X^0$) at 1.99 eV(blue circles) and the optically prepared system that isolates the trion ($X^-$) at 1.94 eV (red squares). The PL intensity of trion is multiplied by 5x to show on the same scale.*

Low, 1-µW, excitation is needed at room temperature to observe only the neutral exciton. Increasing the intensity engages the trion as seen in Figure 2. The PL spectra shown in Fig. 2(a) are from the sample under ambient atmospheric conditions while varying the photoexcitation power. Increasing the excitation amplitude from 23 to 280 µW results in an increase of the trion intensity relative to that of the neutral exciton. Note that while the ratio of PL intensity evolves to favor the trion, the absolute intensity of both the neutral and charged excitons also increase as a function of the excitation power, as expected. Previously, similar behavior was reported which showed the emergence of the trion in the presence of the neutral exciton, however, the intensity of the neutral exciton was always greater than that of the trion [21]. As the excitation power is increased there is a slight red-shift in the trion, but no shift the neutral exciton. We ascribe this phenomenon to degenerate carrier population effects resulting in modification of the band-structure and Fermi level.

In ambient atmosphere, our measurements show the neutral exciton as the dominant peak at low power while the trion is the dominant peak at high excitation power, and this process is completely reversible. This is not the case in vacuum. The PL spectra as a function of increasing excitation power for the sample in vacuum are shown in Fig. 2(b). Once the trion is stabilized by exciting with high power, even after reducing the excitation power to nanowatt levels, the trion emission remains dominant. To reverse this state and regain the neutral exciton emission the sample must be exposed to air. If the sample is exposed to only nitrogen or helium, the neutral exciton will not be recovered. Since PL measures optical emission, this characterizes mostly the lowest energy transition. However, optical absorption-type measurements like our differential reflectance measurement in Fig. 2(c) provides information over a broader spectral region and clearly show (and confirm) the spectral energies of the neutral exciton and trion. This spectrum was taken when the sample was in an intermediate regime, i.e. both the neutral and charged exciton are present. These reflectance measurements allow us to obtain a binding energy of the trion of 33 meV, and have little influence on carrier population effects (unlike photoluminescence). The differential reflectivity has single peaks at the neutral exciton or trion positions when the corresponding PL is at either of these extremes.

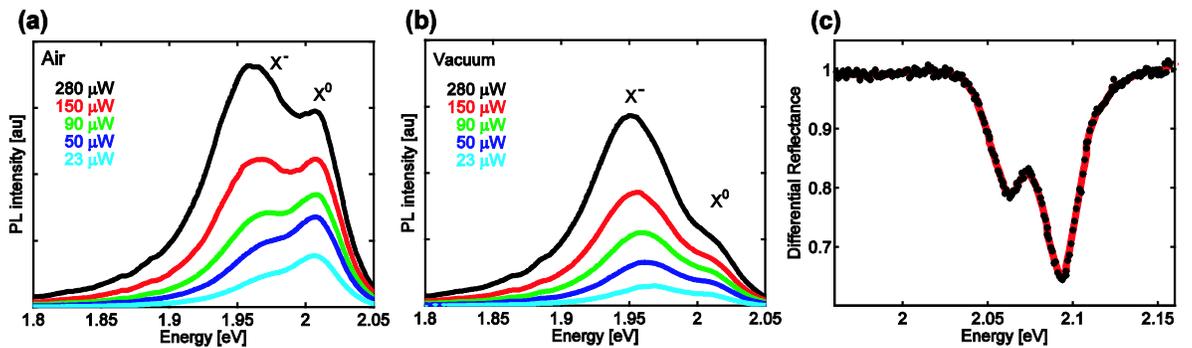

*Figure 2: Photoluminescence of WS2 monolayer to 594nm excitation at room temperature for samples in (a) air and (b) vacuum. The excitation power was increased from 23 to 280 µW in five steps. At low power the PL in air is dominated by the neutral exciton ($X^0$), but the trion ($X^-$) takes over with increasing optical power. In vacuum, the trion ($X^-$) has higher intensity than the neutral exciton ($X^0$) even at low optical power. (c) Differential reflectance at 4K clearly exhibits features for both the exciton and trion.*

To facilitate the transition from the neutral exciton to the trion, we focused a 532-nm laser on the sample to provide a 200-kW/cm$^2$ irradiance. We then raster scanned the laser thoroughly over the entire monolayer region of WS$_2$ (in vacuum). This caused a complete shift in the PL spectrum from the purely neutral exciton PL, as it is in ambient atmosphere, to a purely charged exciton PL. The resulting trion PL is independent of the probing PL excitation power, and further rastering produced no additional change demonstrating the saturation of this process for this optical power level. Thus we are able to create an optically prepared system where either the exciton or trion exists.

After optically preparing this trion system, we further characterize the selected trion state using optical techniques (optical reflectivity, photoluminescence and Raman spectroscopy) at temperatures from 4—320K, and excitation energies from 2.088—2.541 eV (488 – 594nm). The trion characteristics, shown in Fig. 3, can then be analyzed with respect to temperature. The large exciton and trion binding energies in WS$_2$ [16,17] allow us to observe their features even at room temperature, thus, providing analysis over this large temperature range. In Fig. 3(a), the temperature dependence of the PL emission energy fits a standard hyperbolic cotangent relation defined by O'Donnell and Chen [22], and allows us to extract an average phonon energy of 15 meV, in agreement with the acoustic phonons near the K-point in WS$_2$ [23,24]. The temperature dependence of the PL emission linewidth, Fig 3(b), fits well to a Maxwell–Boltzmann distribution of the PL spectrum [25]. Finally, the emission intensity, shown as a function of 1/T in Fig. 3(c), compares well with a standard exponential-decay model [26] since luminescence decreases with increasing temperature due to increased non-radiative transition probabilities at higher temperatures. These measurements match the temperature dependence of an exciton system, verifying our characterization of the optically prepared state as a charged exciton. These measurements also show that we can reliably prepare the initial state of the system at each temperature, thus providing a means for a reproducible surface preparation.

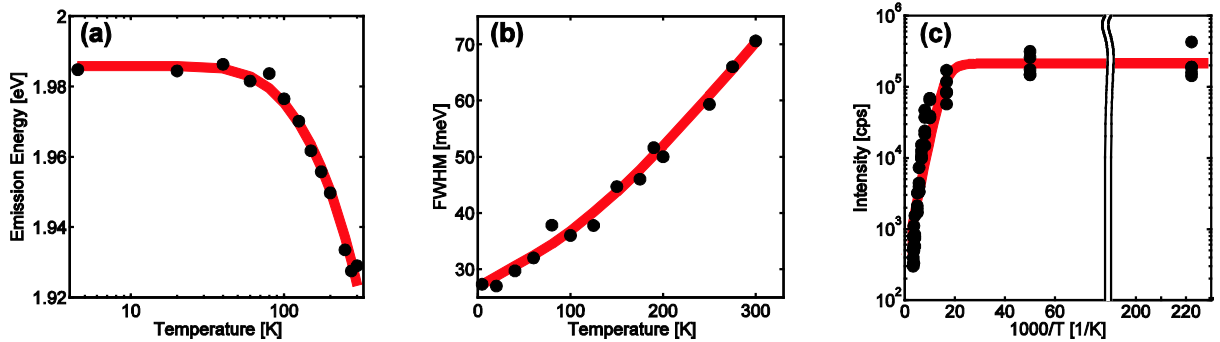

*Figure 3 The trion's (a) emission energy, (b) spectral width, and (c) luminescence intensity as a function of temperature are collected from PL measurements. Data are a compilation from spectra of various excitation energies, and corresponding fits are shown as solid lines through the data.*

While PL and reflectivity explores optical properties relating to the induced dipole moment change in the atomic positions, Raman spectroscopy probes the induced dipole moment due to deformation of the electron shell. We use Raman scattering and observe features consistent with out-of plane $A_{1g}$ and in-plane $E^1_{2g}$ features identified with $WS_2$ [23,24,27–30]. The $A_{1g}$ mode arises from out of plane motion of sulfur atoms, while the $E^1_{2g}$ results from the opposing in-plane motion between the sulfur and tungsten atoms. Fig. 4 shows the Raman spectra of our sample in an ambient environment (solid, blue line), as well as in vacuum after preparing the trion state (dashed, red line). These spectra are normalized to the substrate Si peak. Our ratio of the $A_{1g}$ to the $E^1_{2g}$ is less than one for both cases and confirms our samples are a single $WS_2$ layer [30]. After laser preparation of the sample in vacuum, there are two main differences in the Raman spectra. First, the $A_{1g}$ intensity decreases, as can be seen in Fig. 4(a). This indicates a reduction in the interaction between the sulfur planes, likely arising from surface contaminant desorption.

Second, there is an enhanced feature on the low energy side of the $E^1_{2g}$ Raman peak, see

Figure 4(b). This feature has been identified as a 2LA(M) mode occurring as a second-order mode involving two LA phonons from the M-point in the Brillouin zone [23,24,28,30]. This LA phonon is similar to that seen in Raman scattering from folded acoustic phonons observed in superlattices [31]. While our measurement is configured in a backscatter geometry, the $WS_2$ monolayer is on a $SiO_x$/Si substrate, and therefore some incident light is transmitted through the $WS_2$ and is then reflected from the silicon substrate and returns in a Raman forward-scattering geometry. This produces a mixed forward and backward scattering result and may contribute to the broadened line shape of the 2-LA(M) mode. Indeed such a shift from forward/backward scattering effects has been observed in superlattices [31].

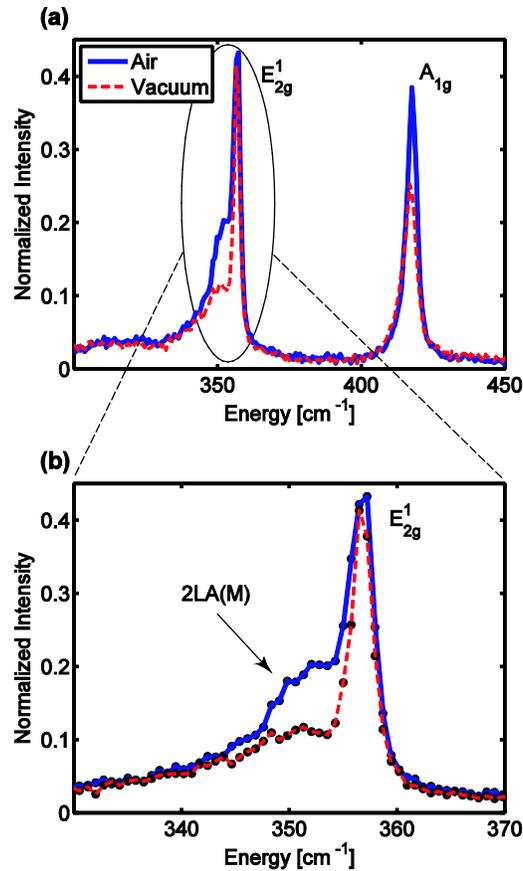

*Figure 4: (a) Raman scattering from sample with (dashed, red line) and without (solid, blue line) optically prepared trion state. The in-plane, $E^1_{2g}$, and out-of-plane $A_{1g}$, peaks are labeled. (b) Magnification of the region near the in-plane $E^1_{2g}$ mode.*

The reduction in this 2-LA(M) mode after preparing the sample may arise from a change in the carrier concentration following the laser preparation. One explanation is dielectric screening from carriers available after surface contaminant desorption. Additionally, this might be explained by trion formation governed by electron localization as suggested by Mak *et al.*[8]. Finally, thiols forming on the $WS_2$ surface may be partially responsible for this effect. Thiols have strong affinity for metal surfaces and rapidly cover these surfaces, however, they quickly degrade in light [32].

Emergence of a trion state was also observed in a study of $WS_2$ by optical pumping with sub-gap (860nm) laser excitation [21]. Comparing this with our data, the most likely mechanism for the optical modification to the trion state is that optical rastering desorbs adsorbates via local heating of the $WS_2$ sample. While local heating is the most likely scenario, adsorbate photo-desorption, photo-doping, or creation of monolayer defects [10] also remain as possibilities. Similar emergence of trion peaks have been observed in other TMDs: $MoS_2$, $MoSe_2$, and $WSe_2$ [10]. As noted by Tongay *et al.*, inhibiting the charge transfer between attached molecules and the TMDs modifies their properties. Molecules such as $O_2$ and $H_2O$ are strongly bound to sulfur-vacancies which dramatically reduces charge density in $MoS_2$ [10]. Subsequent $O_2$ / $H_2O$ removal or charge-transfer inhibition increases the charge density, thereby facilitating our observation of trion creation in $WS_2$ while under vacuum.

In summary, we have demonstrated optical preparation of a monolayer of $WS_2$ such that the photoluminescence is solely from the trion state, suppressing the neutral exciton completely. After isolating the trion state, we observed changes in the Raman spectra: decrease in the $A_{1g}$ intensity and an enhanced feature (attributed to a second-order mode involving two LA phonons from the M-point) on the low energy side of the $E^1_{2g}$ peak. In addition, we can prepare states with varying degrees of neutral and charged excitons, and we observe a 33-meV trion binding energy. This effect can be achieved in seconds, which is much faster than hours-long annealing processes used in previous studies [10]. The luminescence changes appear to be quenched by surface interactions with the ambient environment, evidenced by the fact that the optically prepared state is reversible by exposing the sample to air. This enables the investigation of surface properties for $WS_2$ and other 2D materials. In addition, the optical preparation allows isolated examination of pure trion or exciton states, necessary for observing many physical processes, e.g., time-resolved exciton measurements. This also provides a tool to modify optical emission energy and could potentially modify the band structure in 2D materials enabling applications in optical modulation, tunable optical sources and detectors as well as all-optical sensors. The methodology described here lends itself to optical control of charged exciton states in $WS_2$ (and potentially all TMD and 2D materials) by merely exposing the material to light in controlled environments.

ACKNOWLEDGMENTS

We would like to thank Adam Friedman and Kathy McCreary for sample fabrication as well as Jim Culbertson for assistance with Raman measurements. GK gratefully acknowledges the hospitality and support of the Naval Research Laboratory where the experiments were performed. This work was supported by core programs at NRL and the NRL Nanoscience Institute.